\documentclass[%
 reprint,
superscriptaddress,
 amsmath,amssymb,
 aps,
pra,
]{revtex4-1}

\usepackage[pdftex]{graphicx}
\usepackage{dcolumn}
\usepackage{bm}
\usepackage{amssymb}
\usepackage{amsmath}
\usepackage{color}

\newcommand {\micro}[1]{$\mu$#1}

\definecolor{red}{rgb}{1,0,0}

\begin{document}

\title{Ultra-Broadband Coherent Supercontinuum Frequency Comb}

\author{Axel Ruehl}
\altaffiliation[Present address:~]{Institute for Lasers, Life and Biophotonics, Vrije Universiteit Amsterdam, The Netherlands}
\email{aruehl@few.vu.nl}
\affiliation{IMRA America Inc., 1044 Woodridge Ave., Ann Arbor, MI 48105, USA}

\author{Michael. J. Martin}
\author{Kevin C. Cossel} 
\author{Lisheng Chen} 
\affiliation{JILA, National Institute of Standards and Technology and University of Colorado\\
Department of Physics, 440 UCB, Boulder, CO 80309, USA}

\author{Hugh McKay}
\author{Brian Thomas}
\affiliation{IMRA America Inc., 1044 Woodridge Ave., Ann Arbor, MI 48105, USA}

\author{Craig Benko}
\affiliation{JILA, National Institute of Standards and Technology and University of Colorado\\
Department of Physics, 440 UCB, Boulder, CO 80309, USA}

\author{Liang Dong} 
\altaffiliation[Present address:~]{COMSET/ECE, Clemson University, Clemson, SC 29625}
\affiliation{IMRA America Inc., 1044 Woodridge Ave., Ann Arbor, MI 48105, USA}

\author{John M. Dudley}
\affiliation{Institute FEMTO-ST, CNRS-University of Franche-Comt\'{e} UMR 6174, Besan\c{c}on, France}

\author{Martin E. Fermann} 
\author{Ingmar Hartl }
\affiliation{IMRA America Inc., 1044 Woodridge Ave., Ann Arbor, MI 48105, USA}
	
\author{Jun Ye}
\affiliation{JILA, National Institute of Standards and Technology and University of Colorado\\
Department of Physics, 440 UCB, Boulder, CO 80309, USA}


\date{\today}

\begin{abstract}
\noindent We present detailed studies of the coherence properties of an ultra-broadband super-continuum, enabled by a new approach involving continuous wave laser sources to independently probe both the amplitude and phase noise quadratures across the entire spectrum. The continuum coherently spans more than 1.5 octaves, supporting Hz-level comparison of ultrastable lasers at 698~nm and 1.54~\micro m. We present the first numerical simulation of the accumulated comb coherence in the limit of many pulses, in contrast to the single-pulse level, with systematic experimental verification. The experiment  and numerical simulations reveal the presence of quantum-seeded broadband amplitude noise without phase coherence degradation,  including the discovery of a dependence of the super-continuum coherence on the fiber  fractional Raman gain.   
\end{abstract}



\maketitle


\noindent Highly spatially and temporally coherent super-continua (SC) based optical frequency combs have applications ranging from broadband spectroscopy with high sensitivity and accuracy \cite{coddington2008coherent, *cossel2010analysis}, to the spectral dissemination of optical frequency references, where coherently linking the visible spectrum to the telecom band is crucial for long-haul optical carrier transfer \cite{terra2009phase}. Additionally, and most importantly, highly phase coherent frequency combs directly impact the field of optical frequency standards, where the SC generated with phase-stabilized frequency combs allows comparison of optical frequency standards hundreds of THz apart \cite{foreman2007coherent}. Currently, the thermal noise-limited frequency stability of continuous wave (CW) cavity-stabilized lasers sets a limit to the frequency stability of these optical standards, especially to those based on trapped ensembles of neutral atoms \cite{quessada2003dick}. A direct spectral link between the 1.5~\micro m and visible spectral regions allows novel ultra-stable laser systems operating in the telecom band \cite{kefelian2009ultralow} to be used as the local oscillators for current optical standards. A direct comb-based approach offers more complete spectral coverage and can circumvent frequency doubling of the CW laser or frequency comb, which adds non common-path noise in addition to complexity. 

 \begin{figure*}
 \includegraphics[trim = 10mm 168mm 10mm 43mm, clip, scale=.8]{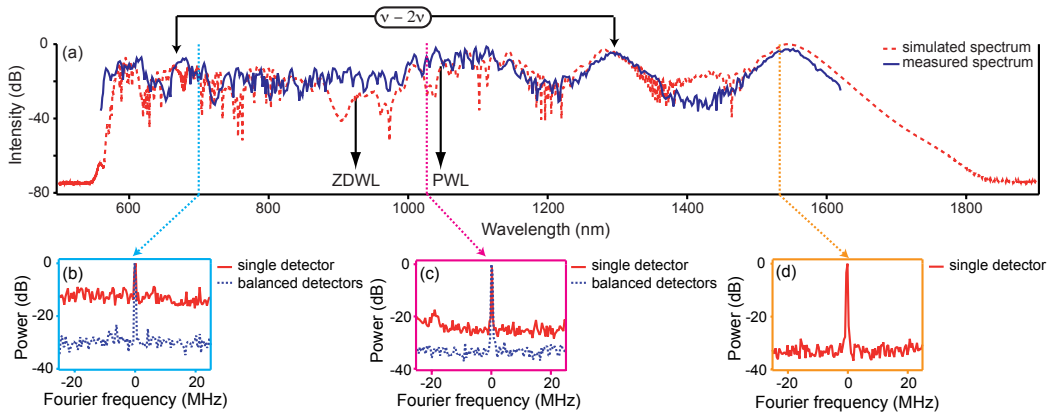}
 \caption{\textbf{(a)} Simulated (dashed) and measured solid super-continuum spectrum at the output of the suspended core fiber. The zero dispersion wavelength (ZDWL) and pump wavelength (PWL) are shown with arrows. The $\nu$--$2 \nu$ interferometer wavelengths used in this work are also indicated. \textbf{(b)} Beat at 698~nm with (dashed) and without (solid) balanced detection. Up to 20 dB suppression of the amplitude noise floor is observed. \textbf{(c)} Beat at 1020~nm with (dashed) and without (solid) balanced detection. \textbf{(d)} Beat at 1.5~\micro m without balanced detection, exhibiting a signal to noise ratio comparable to the balanced cases.  \label{meas_spec}}
 \end{figure*}

In this Letter, we present a new method for verifying supercontinuum optical coherence, which allows us to conclusively demonstrate a spectrum that directly and coherently spans from the visible to the telecom band. This technique, applied to an ultra-broadband spectrum 1.5 octaves wide, utilizes optical heterodyne beat experiments and numerical simulations. This powerful combination has allowed us to identify regions of significant quantum-seeded amplitude noise that are decoupled from the phase noise of the continuum,  proving that the comb phase coherence is preserved across the entirety of the 1.5-octave spectral range. In concert with numerical simulations, we have reached an understanding of the important role that fiber Raman gain plays in maintaining broadband optical coherence in the super-continuum.

The frequency comb  used to generate the supercontinuum is based on a Fabry-Perot-type Yb-similariton oscillator mode-locked with a sub-ps lifetime saturable absorber mirror \cite{harti2006ultra}. While it has been demonstrated that fiber oscillators can produce high quality frequency combs with excellent passive stability \cite{newbury2007low}, pulse durations in the 100 fs range make it challenging to realize super continua with a high degree of coherence over a large spectral region \cite{dudley2006supercontinuum}, as the sensitivity of the broadening process to noise can lead to large fluctuations \cite{nakazawa1998coherence}.  Therefore, we took careful steps to reach near transform-limited nJ-level pulses suitable for SC generation. The 152 MHz pulse train was amplified in a Yb-doped double-clad phosphosilicate fiber with 12~\micro m core diameter pumped at 976 nm. To avoid self-phase modulation and hence to minimize amplitude-to-phase noise conversion in the fiber amplifier, a linear chirped-pulse amplification scheme was chosen, and it has been demonstrated that this amplification scheme is compatible with low-noise phase control \cite{schibli2008optical, *ruehl201080}. With additional fiber-based compensation for higher-order dispersion contributions, the pulses could be recompressed to $< 80$~fs with a grating compressor as characterized by frequency resolved optical gating. We coupled the 1.2~nJ-pulse train into a highly nonlinear suspended core fiber \cite{Fu2008} with a zero dispersion wavelength tuned to approximately 950~nm in order to optimize the spectrum, which spanned from 570~nm up to 1.66~\micro m at the -30 dB-level as shown in Fig.~\ref{meas_spec}~(a). We note that the measured spectrum exhibits excellent agreement with the simulation (dashed line in Fig.~\ref{meas_spec}~(a)), which we later discuss in detail.  

In order to test the phase coherence of the supercontinuum in different spectral regions, we performed free-running beat experiments with several CW lasers. Amplitude noise was identified with the help of dual-balanced detection, which cancels out common mode noise \cite{yuen1983noise}. We implemented this scheme to obtain beat notes at 698~nm and 1020~nm, respectively (shown in Fig.~\ref{meas_spec}) with signal-to-noise ratio (SNR) of $> 30$~dB at 100~kHz resolution bandwidth (RBW). As shown in Figs. \ref{meas_spec}~(b) and \ref{meas_spec}~(c), the amplitude noise floor could be suppressed by up to 20~dB at 698~nm and 8~dB at 1020~nm by using dual-balanced detection compared to the measurement with a single detector. This observation supports the discovery of broadband amplitude noise of the comb without the presence of phase noise in these spectral regions. In other spectral regions (e.g. at 1540~nm) we observed that dual-balanced detection was not necessary to obtain a similar SNR (shown in Fig.~\ref{meas_spec}~(d)). More than 30~dB SNR was also obtained for beat signals at 771~nm and for the $\nu-2 \nu$ heterodyne beat signal representing the carrier-envelope offset frequency, $f_{0}$. Self-referencing the comb was possible with  $c/ \nu = 1150$~nm to $c/ \nu = 1300$~nm (and the corresponding second harmonic, $2\nu$) providing the freedom to detect other beat notes in this region. 

\begin{figure}
 \includegraphics[trim = 0mm 1mm 0mm 0mm, clip, width=7.5 cm]{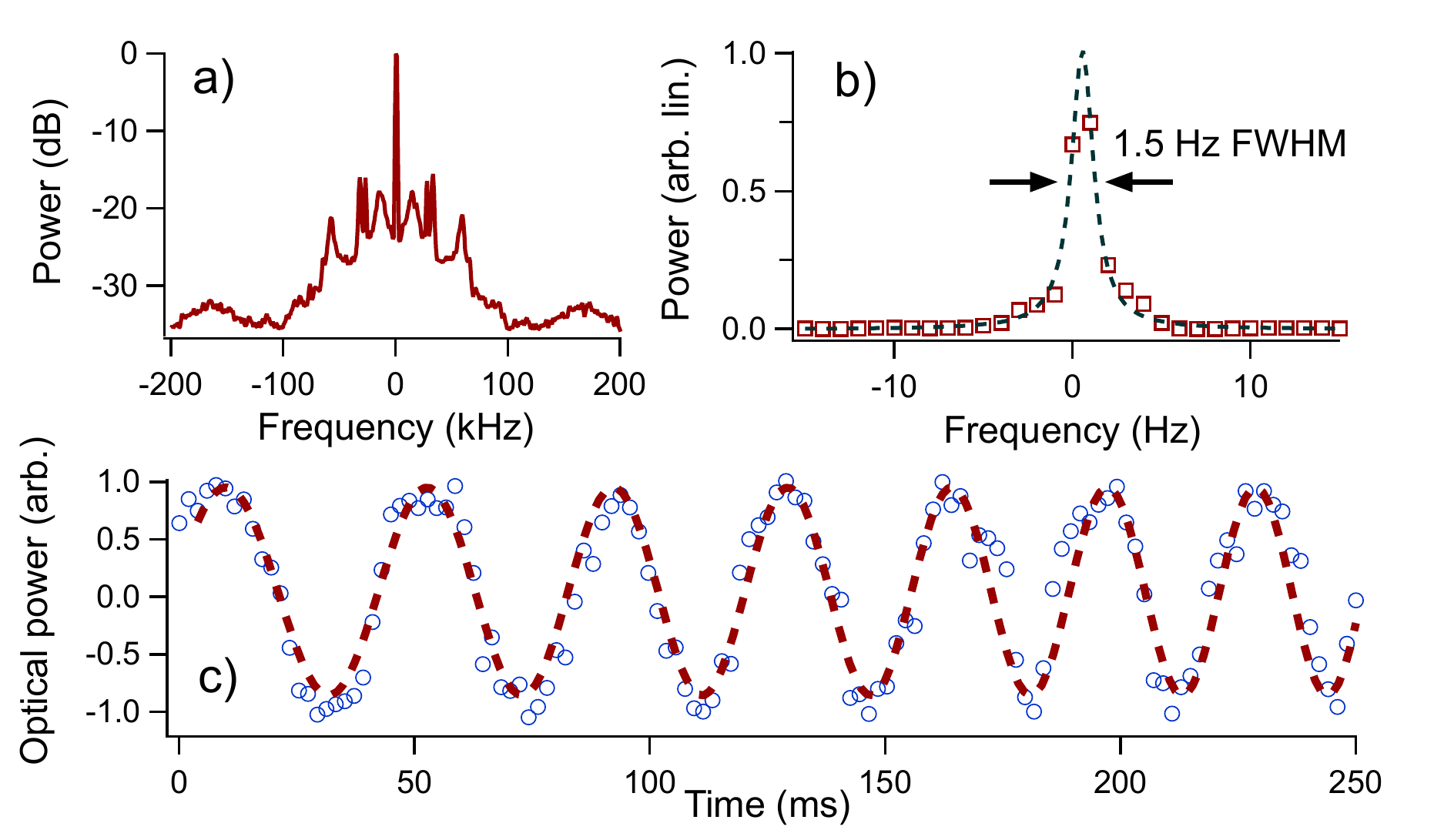}
 \caption{Optical heterodyne beat results at 1.54~\micro m attained via comb-facilitated comparison of ultrastable lasers at 698~nm and 1.54~\micro m. \textbf{(a)} Broad (1 kHz RBW) and \textbf{(b)} narrow (1 Hz RBW) view of the resulting heterodyne beat at 1.54~\micro m (squares) and Lorenztian fit (dashed line). \textbf{(c)} Time domain representation of the heterodyne beat (circles) and fit (dashed line), demonstrating a coherence time $> 200$~ms. The fitting function is a chirped sine wave to account for relative linear drift between the two ultrastable lasers.\label{OL}}
 \end{figure}

The free running beat signals demonstrate that the SC is coherent beyond the 1~\micro s time scale, permitting coherent beats with high SNR to be observed across the entire spectrum. We additionally performed a precision test of the comb coherence between two distant spectral regions to probe the coherence at the 1 s time scale and demonstrate an important technological milestone: the direct and precise comparison of ultrastable lasers in the visible and telecom bands. The self-referenced frequency comb was locked to a sub-Hz linewidth optical local oscillator at 698~nm used in the JILA Sr optical lattice clock \cite{ludlow2007compact}. Then the fully stabilized comb was compared to a second stabilized CW laser system at 1.54~\micro m based on a compact and energy-efficient diode laser (Redfern Integrated Optics) with a free-running linewidth of $< 10$~kHz. The optical power of the diode laser was enhanced with an low noise Er-doped fiber amplifier capable of powers up to 300~mW. The laser was stabilized by injection current control to an ultra-high finesse optical cavity based on a ultra-low expansion (ULE) glass spacer and fused silica mirrors, with finesse of 180,000 and a linewidth of 23~kHz, using the Pound-Drever-Hall scheme.  The simple and robust lock  facilitated unattended operation over many days with Hz-level short-term stability. The stabilized 1.54~\micro m laser light was delivered to the frequency comb via a noise-cancelled fiber. The corresponding beat-signal between the frequency comb and the stabilized 1.54~\micro m CW laser measured at 100~kHz RBW is shown in Fig.~\ref{OL}~(a). The beat signal was further mixed down to 50 kHz and measured with a Fourier-analyzer with 1~Hz RBW. From this measurement, one can deduce a coherence time of  $>200$~ms corresponding to $> 3 \times 10^{7}$ pulses from our 152~MHz pulse train. A similar value was also found by fitting the time record of the out-of-loop beat note. As depicted in Fig.~\ref{OL}~(c), a chirped sinusoidal fitting function of characteristic time span of $> 200$~ms could be successfully applied to the experimental data. 
 
To better understand the spectral broadening process and SC coherence properties, we performed numerical simulations of the pulse dynamics within the nonlinear fiber. The propagation of the radially averaged electric field envelope, $A\left(z,t\right)$, is described by a generalized nonlinear Schr\"{o}dinger equation \cite{blow1989theoretical, dudley2006supercontinuum},
\begin{align}
&\frac{\partial A\left(z,t\right)}{\partial z}  - i\mathcal{F}^{-1}\Big[\big(\beta\left(\omega\right) - \omega \beta_{1} - \beta_{0} \big) \tilde{A}\left(z,\omega \right) \Big]= \nonumber \\
&\! \!  i \gamma \left(1 + \frac{i}{\omega_{0}} \frac{\partial}{\partial t}\right)  \left[A\left(z,t\right) \int R\left(t'\right) \left| A\left(z, t-t'\right)\right|^{2} dt'\right]. 
\end{align}
The terms on the left side of the equation describe the contributions of the fiber dispersion to all orders while the terms on the right side take into account the relevant $\chi^{(3)}$ effects, namely Raman scattering and the Kerr effect. Specifically, $\tilde{A}\left(z,\omega \right)$ is the Fourier transform of $A\left(z,t \right)$, $\mathcal{F}\left[A\left(z,t \right)\right]$; $\beta_{k} \equiv \frac{\partial^{k}}{\partial \omega^{k}} \beta\left(\omega\right) \big|_{\omega_{0}}$, where $\beta\left(\omega\right)$ is the propagation constant and $\omega_{0}$ is the carrier frequency; $\gamma$ is the nonlinear coefficient and is proportional to $n_{2}$, the nonlinear refractive index; and $R\left(t\right)$ is given by $R\left(t\right)  = \left(1-f_{R} \right)\delta \left(t\right) + f_{R} h_{R} \left(t\right)$. The Raman fraction, $f_{R}$, sets the ratio of Kerr to Raman non-linearity, $h_{R} \left(t\right)$ is the time-domain Raman response \cite{blow1989theoretical}, and $\delta \left(t\right)$ is the Dirac delta function. 

We measured the pulse energy coupled into the fiber (1.2~nJ), input spectrum, and fiber dispersion and used these as the basis for our simulation parameters. The input pulses were modeled as unchirped hyperbolic secant pulses of 80~fs duration (full width at half maximum) with a center wavelength of 1050 nm. The fiber had a nonlinearity ($\gamma$) of approximately 40~W$^{-1}$km$^{-1}$ and also an increased $f_R$, due to germanium doping, from the standard value of 0.18 for fused silica. We found that the simulated spectrum best matched the measured spectrum for $f_{R} = 0.28$ (Fig.~\ref{meas_spec}~(a)), consistent with the Raman gain increase observed in \cite{oguama2005simultaneous}. The injected soliton order \cite{dudley2006supercontinuum}, $N \simeq \sqrt{(0.5 \gamma E_0 T_0)/|\beta_2|}$ where E$_0$ is the pulse energy and T$_0$ is the pulse duration, was 19.2.

\begin{figure*}
 \includegraphics[trim = 10mm 179mm 10mm 38mm, clip, scale=.7]{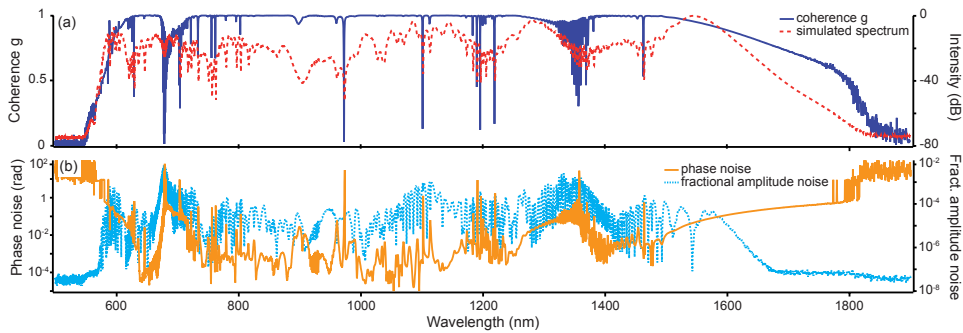}
 \caption{Simulation results for the optical spectrum and coherence properties after broadening in the suspended core fiber. \textbf{(a)} Simulated spectrum (dashed, right) and calculated quantum noise-limited pulse to pulse coherence (solid, left). \textbf{(b)} Contributions of amplitude (dashed,right) and phase (solid, left) noise to the total coherence function. In many spectral regions, the presence of amplitude noise does not presage the presence of phase noise, indicating that degradation of $g\left(\omega\right)$ is not necessarily indicative of a loss of phase coherence.  \label{Simul}}
 \end{figure*}

In order to model quantum-limited SC coherence, we performed multiple simulations with a quantum shot noise seed perturbing the input pulse,  $A \left(t\right)$, such that  $A \left(t\right) \rightarrow A \left(t\right) + \Delta A \left(t\right)$. In this case, where the square modulus of $A$ represents optical power, $\Delta A \left(t\right)$ is described stochastically by $\langle \Delta A^{\ast}\left(t\right) \Delta A \left(t+\tau\right) \rangle = (h \nu/2) \delta \left(\tau\right)$ \cite{corwin2003fundamental, paschotta2004noise}, and we treat it as a Gaussian-distributed random variable in both quadratures. We also include the effects of spontaneous Raman scattering along the fiber according to the procedure of \cite{corwin2003fundamental}. The first-order coherence, $g\left(\omega\right)$, defined as
\begin{equation}
g\left( \omega \right) = \frac{\left| \langle \tilde{A}_{i} \left(\omega\right) \tilde{A}^{\ast}_{j} \left(\omega\right) \rangle_{i \neq j} \right|}{\sqrt{\langle \left|  \tilde{A}_{i} \left(\omega\right) \right|^{2} \rangle \langle \left| \tilde{A}_{j} \left(\omega\right) \right|^{2} \rangle }},
\end{equation}
was calculated numerically by an ensemble average over 100 independently simulated spectra according to Ref. \cite{dudley2002numerical}, in 50~GHz spectral bins.  Fig.~\ref{Simul}~(a) shows $g\left(\omega\right)$ along with the calculated spectrum, and it can be seen that there are significant drops in the coherence in isolated spectral regions that still posses significant optical power. Another interesting note is that the coherence degrades smoothly across the first Raman soliton (with a corresponding increase in phase noise).

Numerically, we identified several parameters responsible for a significant loss of coherence. In particular, our quadrature-sensitive technique for measuring supercontinuum coherence led us to discover a dependence on the fractional Raman contribution ($f_R$) that has not been studied previously. We calculated the average coherence of the longest wavelength (first ejected) soliton and saw a decrease in coherence from 0.9 to 0.7 for $f_R$ varying from 0.28 to 0.2 with the soliton center wavelength fixed by increasing the fiber length. The loss of coherence was most dramatic on the long-wavelength side of the soliton. We also noticed reduced coherence with increasing pulse energy and with increasing pulse duration (both increasing the soliton order) as has been observed in other studies \cite{demircan2007analysis}; however, we see a coupling between these parameters and $f_R$ that has not been noted previously. With a higher value of $f_R$, the coherence was significantly less sensitive to pulse duration (slight loss of coherence in the soliton at 120 fs) and energy than with low values of $f_R$, where we saw strong loss of coherence in the first soliton and also low coherence across the spectrum with 90 fs pulses. These results are significant in that they demonstrate that intrapulse stimulated Raman scattering plays a central role in generating supercontinuum spectra with high coherence. Specifically, although competition between noise amplification through modulation instability and coherent spectral broadening through soliton fission has already been identified as a key factor determining supercontinuum coherence, our results suggest that a higher fractional Raman contribution can mitigate the addition of noise to the propagating solitons. The increased rate at which the ejected solitons shift to longer wavelength reduces the detrimental effect of the modulation instability gain on the soliton amplitude and phase. Designing and fabricating fibers with higher fractional Raman contribution (e.g., higher Ge-doping) may therefore represent an important technological tool for generating low noise supercontinua, especially in cases where source parameters (pulse duration, power) cannot be independently controlled.


To better elucidate the experimental observations, we consider the different contributions to $g\left(\omega\right)$ arising from amplitude and phase noise separately (\textit{i.e.}, calculating the variance of the argument and amplitude of $\tilde{A} \left(\omega\right)$). As evident in Fig.~\ref{Simul}~(b), it was numerically found that certain spectral regions feature decreased coherence arising from amplitude noise but exhibit only negligible phase noise contributions. The amplitude noise observed in the different spectral regions qualitatively agrees with the experimental trend observed in Fig.~\ref{meas_spec}. Interestingly, and in relation to the previous discussion, the simulations show that the loss of coherence of the first soliton is due primarily to phase noise. This distinction between amplitude and phase contributions is in contrast to previous experimental studies, where SC coherence was demonstrated by the use of a modified Young experiment with two sources or delayed pulse interferometry, where such a clear distinction could not be made \cite{bellini2000phase, Lu2004}. Furthermore, direct comparison between 1.54~\micro m and 698~nm CW lasers demonstrates the capability of delivering Hz-level (1~s-level) optical linewidth (coherence time) across the whole optical spectrum and hence the connection of optical local oscillators at 1.5~\micro m to optical frequency standards. Supporting this experimental observation, numerical simulations of up to 1000 pulses show that the total phase standard deviation does not change, indicating that the quantum noise-induced phase deviation is bound at some level and is not a random walk, which would degrade relative linewidth. This is consistent with other tests of comb coherence which reach the same conclusion \cite{martin2009testing}. SC generation can thus be seen as a white phase noise process without an accumulative component. This further confirms that the observed linewidth in the out-of-loop comparison was not compromised by the SC. 

In conclusion, we have provided a systematic understanding of the coherence properties of a 1.5 octave wide SC generated from an amplified Yb-fiber frequency comb via a new quadrature-sensitive technique utilizing CW laser sources. Contributions arising from amplitude and phase noise were numerically calculated and experimentally confirmed. It was demonstrated that spectral regions with increased amplitude noise can still be utilized for high quality optical phase measurements when dual-balanced detection is employed. We identified relevant fiber parameters, crucially the fractional Raman contribution and its effect on soliton shifting rate, which will allow further improvement in the design of highly nonlinear fibers for optimized SC coherence properties. Importantly, we demonstrated coherence transfer over more than one octave between a ULE cavity stabilized diode-lasers at 1.54~\micro m laser and 698~nm representing the largest spectral gap directly spanned between two ultra-stable lasers by a frequency comb to date. Limited by the linewidth of the 1.54~\micro m laser, a coherence time of $> 200$~ms was found. This technique could enable more detailed  and quantitative studies of  SC coherence.

\begin{acknowledgments}
We acknowledge helpful discussions with R. Holzwarth and T. Wilken. The work is funded by NIST and NSF. L. Chen is currently at WIPM, China. A. Ruehl is supported by the Alexander von Humboldt Foundation.
\end{acknowledgments}


%

\end{document}